\documentstyle[twocolumn,epsbox]{jpsj}

\def\mbf(#1){\mbox{\boldmath $#1$}}
\def\runtitle{Study on nonmagnetic impurities in $t$-$J$ model}
\def\runauthor{Hideki {\sc Tanaka}, Kazuhiro {\sc Kuboki} 
               and Manfred {\sc Sigrist}}

\title
{Study on nonmagnetic impurities in the superconducting state 
of two-dimensional $t$-$J$ model}

\author{Hideki Tanaka$^a$, Kazuhiro Kuboki$^a$ and Manfred Sigrist$^b$}

\inst{$^a$Department of Physics, Kobe University, Kobe 657-8501, Japan \\
$^b$ Yukawa Institute for Theoretical Physics, Kyoto University, 
           Kyoto 606-8502, Japan}

\recdate{\today}

\abst{  
We study the effect of nonmagnetic impurities in the superconducting state 
of the $t-J$ model. The Bogoliubov de Gennes equation is derived using 
a slave-boson mean-field approximation, and we solve it numerically 
at $T=0$. The $d_{x^2-y^2}$-wave order parameter is 
suppressed near an impurity and its spatial variation induces the 
extended $s$-wave component. 
We also find the local charge density wave (CDW)-like feature due to 
the Friedel oscillations. 
This leads to the splitting of the expected zero-energy Andreev bound
state. Therefore the system is unlikely introducing a localized state of other
symmetry such as superconductivity with broken time reversal symmetry
or antiferromagnetism.
}

\kword{High-Tc superconductivity, t-J model, d-wave symmetry, 
nonmagnetic impurity}

\begin{document}
\sloppy
\maketitle

The symmetry of the superconducting (SC) state in high-$T_c$ 
superconductors (HTSC) has been a subject of intensive study 
since it is an
important clue to clarify the mechanism of their 
superconductivity.\cite{Scl}
Now it is established that the SC state has a predominantly 
$d_{x^2-y^2}$-wave character with a possible mixture of an $s$-wave 
component due to the orthorhombic lattice distortion in some systems.  
In $d$-wave superconductors the effect of nonmagnetic impurities 
is quite different from that of conventional ($s$-wave) superconductors 
because of the phase structure of the pair wavefunction and
the resulting presence of nodes in the excitation gap. This
problem has been studied intensively by many authors, mostly based on
weak coupling theories \cite{Imp1,Imp2,Imp3,bala,Imp4}.
In this letter we examine the effect of nonmagnetic impurities in the 
SC state of the $t$-$J$ model on a square lattice. 
This model describes the low-energy electronic properties of 
HTSC\cite{PWA,ZR} where strong correlation effects for the electrons 
are known to be important. 
Mean-field (MF) theories based on a slave-boson method predict a 
superconducting state with a $d_{x^2-y^2}$-symmetry.\cite{SHF,KL} 
They may also explain the magnetic\cite{TKF,FKNT} as well as 
the transport\cite{NL} properties of HTSC 
if the gauge fields representing the fluctuations around the MF 
solutions are taken into account. 
Therefore the effect of nonmagnetic impurities in the SC states of the 
$t$-$J$ model is of particular interest.\cite{Matsu,Tsuchi} 

We treat the $t$-$J$ model with the Hamiltonian including 
that of the nonmagnetic impurities
\begin{equation} 
  \begin{array}{rl}
  H  =& - \displaystyle t \sum_{<i,j> \sigma} 
    ( {\tilde c}_{i,\sigma}^{\dagger} 
        {\tilde c}_{j,\sigma} + h.c.) \nonumber \\ 
     +& \displaystyle J \sum_{<i,j>} {\mib S}_i \cdot {\mib S}_j 
     + V_0 \sum_{l\sigma} {\tilde c}_{l,\sigma}^{\dagger} 
       {\tilde c}_{l,\sigma}  \,
\end{array}
\end{equation}
where $\langle i,j\rangle$ and $l$ denote nearest-neighbor bonds and the 
impurity sites, respectively, 
and ${\tilde c}_{i\sigma}$ is an electron operator within the Hilbert space 
excluding double occupancy. Here ${\mib S}_i$ is the spin-1/2 operator at a
site $i$ 
and $J (> 0)$ is the antiferromagnetic superexchange interaction.   
The impurity potential $V_0$ is taken to be $V_0 \gg J$.   
We use the slave-boson method to enforce the condition of no 
double occupancy by introducing spinons ($f_{i\sigma}$; fermion) 
and holons ($b_i$; boson) (${\tilde c}_{i\sigma} = 
b_i^\dagger f_{i\sigma}$). 
We decouple this Hamiltonian by a mean-field approximation (MFA).
In the following we consider only the case of $T=0$, so that holons are 
Bose condensed. Then the mean-field Hamiltonian is written 
in terms of spinons only, 
\begin{equation}
{\cal H}_{MFA} = \sum_i\sum_j [f_{i\uparrow}^\dagger, f_{i\downarrow}]
\left [\begin{array}{clcr}
W_{ij} & F_{ij} \\
F_{ji}^* & -W_{ji}
\end{array}\right ]
\left [\begin{array}{clcr}
f_{j\uparrow} \\
f_{j\downarrow}^\dagger
\end{array}\right ]
\end{equation} 
with  
\begin{equation}
\begin{array}{rl}
W_{ij} =& \displaystyle -\big(t\delta + \frac{3}{4}J\chi_{ij}\big) 
           \sum_\mu \delta_{j,i+\mu} - \lambda \delta_{ij} 
           + V_0\delta_{il}\delta_{jl} \\
F_{ij} =& \displaystyle -\frac{3}{4}J\Delta_{ij}\sum_\mu \delta_{j,i+\mu}
\end{array}
\end{equation}
where $\delta$ and $\lambda$ are the doping rate and the chemical potential, 
respectively,  and $\mu = \pm {\hat x}, \pm {\hat y}$. The summations for  
$i$ and $j$ are taken over all sites. 
Here the singlet RVB order parameter (OP), $\Delta_{ij}$, and the bond OP, 
$\chi_{ij}$, are defined as 
\begin{equation}
\Delta_{ij} = 
\langle f_{i\downarrow}f_{j\uparrow}-f_{i\uparrow}f_{j\downarrow}\rangle/2, 
\ \ 
\chi_{ij} = \sum_\sigma \langle f_{i\sigma}^\dagger f_{j\sigma}\rangle 
\end{equation}  
respectively, for each nearest neighbor bond. 
We consider the solution where $\Delta_{ij}$ may be complex, while 
$\chi_{ij}$ is taken to be real (i.e., $\chi_{ij} = \chi_{ji}$). 
We do not consider the possible antiferrromagnetic correlation here. 

Without impurities ${\cal H}_{MFA}$ can be easily diagonalized by the 
Bogoliubov transformation assuming spatially uniform $\chi$ and $\Delta$.
In the presence of inhomogeneities, however, we have to take  
their spatial variations into account. They are most conveniently 
described by the Bogoliubov de Gennes (BdG) equation.\cite{dG} From 
eq.(2) it is obtained as 
\begin{equation}
\epsilon_n 
\left [\begin{array}{clcr} u_n(i) \\ v_n(i) \end{array}\right ]
= \sum_j
\left [\begin{array}{clcr}
W_{ij} & F_{ij} \\
F_{ji}^* & -W_{ji}
\end{array}\right ]
\left [\begin{array}{clcr} u_n(j) \\ v_n(j) \end{array}\right ]. 
\end{equation}
Here ($u_n(i), v_n(i)$) is the wave function at a site $i$, and $\epsilon_n$ 
is the corresponding energy eigenvalue. 
The OP's $\chi_{ij}$ and $\Delta_{ij}$, and the 
doping rate $\delta$ (we take $\lambda$ as an input parameter) 
are expressed in terms of $u_n(i)$ and $v_n(i)$,
\begin{equation}
\begin{array}{rl}
\chi_{ij} =& \displaystyle 2\sum_n \big[ u_n^*(i)u_n(j)f(\epsilon_n) 
           + v_n(i)v_n^*(j)\big\{1-f(\epsilon_n)\big\}\big] \\
\Delta_{ij} =& \displaystyle \frac{1}{2} \sum_n 
  \big[ \big\{v_n^*(i)u_n(j)+v_n^*(j)u_n(i)\big\}f(\epsilon_n) \\   
 & \displaystyle \hskip0.5truecm - \big\{u_n(i)v_n^*(j)+u_n(j)v_n^*(i)\big\} 
   \big\{1-f(\epsilon_n)\big\}\big] \\
& \\
\delta =& \displaystyle 1 - \frac{2}{N}\sum_n \sum_i
\big[|u_n(i)|^2f(\epsilon_n) \\ 
 & \hskip2.0truecm  + \displaystyle |v_n(i)|^2\big\{1-f(\epsilon_n)\big\}\big]
\end{array}
\end{equation} 
where $f(\epsilon)$ is the Fermi distribution function and $N$ is the 
total number of lattice sites.  

We numerically solve the self-consistency equations (5) and (6).
For simplicity we assume that $\chi_{ij}$ has a spatially uniform value 
(obtained by MFA without impurity) except for the bonds connected to the 
impurity sites where $\chi_{ij}=0$. 
First we choose $\lambda$, and assume 
some initial values of $\Delta_{ij}$ and $\delta$. 
Inserting these into the BdG equation 
we diagonalize the resulting matrix 
using Householder method (the size of the matrix is 
$2M^2\times 2M^2 (= 2N\times 2N)$, if the system size is $M\times M
(=N)$).\cite{Nishi}  
Then we recalculate $\Delta_{ij}$ and $\delta$ according to eq.(6). 
This procedure is iterated until the convergence is reached. From 
$\Delta_{ij}$ we define a $d_{x^2-y^2}$- and an  
extended $s$- wave OP component on the site $ i $, 
\begin{equation}
\begin{array}{rl}
\Delta_d(i) \equiv& \displaystyle (\Delta_{i,i+x}+\Delta_{i,i-x}-
\Delta_{i,i+y}-\Delta_{i,i-y})/4 \\
\Delta_s(i) \equiv& \displaystyle (\Delta_{i,i+x}+\Delta_{i,i-x}+
\Delta_{i,i+y}+\Delta_{i,i-y})/4,
\end{array}
\end{equation}
where $ x $ and $ y$ correspond to one lattice constant in $x$- and
$y$-direction, respectively.
We have checked that the values obtained for the system without impurities
agree well with those in the usual mean-field calculation, 
when $M \geq 16$. In this case only $\Delta_d$ is finite and it has a 
spatially uniform value. 

Now we turn to the results for the system with an impurity. 
We consider the case where the concentration of impurities is low so that 
the states around the impurities can be treated independently. 
We fix $t/J =3$ throughout in the following, and the doping rates are chosen 
to be $\delta=0.20$. 
(We have studied the system with $0.05 \leq \delta \leq 0.20$.
The results are qualitatively the same for all cases.)
In Fig.1 the spatial variation of $\Delta_d$ is shown, which 
is suppressed near the impurity, and the effect 
is strongest along the diagonals of the square lattice. 
This is due to the interference effects for momenta close to the gap
nodes (sign change of the pair wave function).
The extended $s$-wave component is induced in the region where $\Delta_d$ 
is not uniform (Fig.2), and  $\Delta_s$ at sites rotated 90 degree around the 
impurity have the same magnitude but the opposite sign. 
It vanishes for sites located along the diagonals passing 
through the impurity site. 
($\Delta_d$ and $\Delta_s$ are always real in the solutions we found.)

We analyze the above results qualitatively by using the 
GL theory (though GL theory is not quantitatively valid at $T=0$). 
The GL free energy for the system with an impurity is written 
generically as\cite{SigUe}
\begin{equation}
\begin{array}{rl}
{\cal F} = & \displaystyle \int d^2 {\mib r} \big[\sum_{j=d,s} 
\{ \tilde{a}_j(T) |\Delta_j|^2 
+ \beta_d |\Delta_j|^4 + K_j |{\mib \nabla} \Delta_j|^2 \} \\  
& \\
&    + \gamma_1 |\Delta_d|^2 |\Delta_s|^2 + \frac{1}{2} \gamma_2
(\Delta_d^{*2} \Delta_s^2 + \Delta_d^2 \Delta_s^{*2}) \\ 
& \\
& + \tilde{K} \{ (\partial_x \Delta_d)^* (\partial_x \Delta_s) - 
(\partial_y \Delta_d)^* (\partial_y\Delta_s) + c.c \} \\ 
& \\
& \displaystyle + g_d \delta({\mib r})|\Delta_d|^2 
+ g_s \delta({\mib r})|\Delta_s|^2 \big]
\end{array}
\end{equation}
where the last line represents the effect of the impurity located at 
${\mib r}=0$ ($g_d > 0$, $g_s > 0$), and the crystal axis directions are 
denoted as $x$ and $y$.
We note that the coupling term like 
$(\Delta_d\Delta_s^* + c.c.)$ should not arise due to the symmetry.
Due to the $g_d$ term $\Delta_d$ is suppressed and its gradient becomes 
finite over the range of the coherence length. 
Then $\Delta_s$ is induced through the mixed gradient (${\tilde K}$)-term.
In the ${\tilde K}$-term the gradients in $x$ and $y$ directions have 
opposite sign following the $ \Delta_d $-symmetry, and the induced
$\Delta_s$ must reflect this property.  
Therefore $\Delta_s$ should change sign 
under 90-degree rotation around the impurity.
When the continuous change of the phases of $\Delta_d$ and $\Delta_s$ were 
allowed, $\Delta_s$ could change sign 
under 90-degree rotation with keeping its amplitude finite 
(i.e., $\phi_{ds} = \pm \pi/2$ along the diagonals). 
If the relative phase between $\Delta_d$ and $\Delta_s$ ($\phi_{ds}$) is 
neither 0 or $\pi$, 
the state has $(d+is)$- or $(d-is)$-symmetry and breaks 
time reversal symmetry ${\cal T}$.\cite{TBrk1,TBrk2} 
In this case the gap nodes disappear in the vicinity of the impurity
and the system could gain more condensation energy. 
In our calculation no solution of this type with $\Delta_d$ and
$\Delta_s$ in a complex combination appeared for the 
doping range $0.05 \leq \delta \leq 0.20$.  
In the following we would like to discuss a possible reason for this
result. 

It has been discussed previously based on the T-matrix approximation
that non-magnetic impurities in a $d$-wave superconductor create a
subgap bound state \cite{bala}, similar to Shiba's bound state in a conventional
superconductor around a magnetic impurity \cite{shiba}. In the T-matrix
formulation 
it is found that a single bound state occurs at the Fermi level
(zero-energy) in the unitary limit, i.e. for a strong impurity
potential \cite{bala}. This is understood also within the BdG formulation as an
Andreev bound state \cite{ZBS1}.
Because a strong scattering center destroys the
gap in the near vicinity of the impurity, 
quasiparticles can be
trapped in this potential well and is subject to  Andreev reflection
at the walls of this well. Their energies are determined by the
phase of the gap function in the momentum directions which the
particle-hole trajectories connect via scattering at the impurity \cite{ZBS1}.
For the d-wave superconductor there are two possible phase differences, 
0 and $ \pi $. The former leads to a state at the bulk gap
value while the later generates a zero-energy bound state similar to
the one at the [110]-oriented surface of a $ d_{x^2-y^2} $-wave
superconductor \cite{ZBS1,ZBS2,ZBS3,ZBS4,ZBS5,ZBS6,ZBS7}. The presence
of the latter bound states should yield 
a finite contribution of local density of states (LDOS) at the
Fermi energy. In this case it could be energetically favorable to
twist the OP phase in order to shift the zero-energy bound state 
away from the Fermi energy. 
Such a state would then correspond to the state with locally broken
time reversal symmetry around the impurity as mentioned above. 
This effect is, of course, a matter of competition between the energy
gain of the quasiparticles and the energy expense due to phase
gradients (yielding supercurrents) introduced by the twist.  

In order to examine this view we consider the LDOS which is denoted as
$N({\mib r}_i,\omega)$ on the site $ i $,

\begin{equation}
N({\mib r}_i, \omega) = \displaystyle \sum_n 
\big[|u_n(i)|^2 \delta(\omega-\epsilon_n) + 
|v_n(i)|^2 \delta(\omega+\epsilon_n)\big].
\end{equation}
In Fig.3 we show the result for $N({\mib r}_i, \omega)$ for a
nearest neighbor site of the  
impurity ($\delta = 0.20$). At an energy much lower than the gap value 
($ \Delta_d \sim 0.12 $) we find bound states. The bound state energy
is different from the naively expected value, i.e., zero, leading to a small
gap between the state below and above the Fermi energy. The LDOS is
different for these two states, since the doped system is not
particle-hole symmetric.

The splitting of the expected zero-energy bound state into levels
slightly above and below zero occurs obviously without the $ {\cal T}
$-violation, since our solution has only real superconducting
OP's. This splitting may have various reasons. 
The validity of arguments based on the
Andreev approximation might not be guaranteed here. Another important
aspect is, however, connected with charge density oscillations as
shown in Fig.4.  They originate from the Friedel oscillations due
to the presence of the impurity. The dominant $ Q $-vector lies along the
[110]-direction which is for $ \delta = 0.2 $ very close to $ {\bf Q}
= (\pi,\pi) $ and leads to a nearly commensurate charge density
staggering. In the vicinity of the impurity these lead to a charge density 
wave (CDW) 
like environment with an effective ``doubling of the unit cell'' so that
the local opening of a gap is expected. We would like to emphasize, however, 
that this CDW feature is not the result of an instability triggered by 
the quasiparticle density of states at the Fermi level, but is driven by the
presence of the impurity. 
The fact that there is no quasiparticle peak in the LDOS at zero
energy leads to the
conclusion that a transition to a state with local $ {\cal T}
$-violation or antiferromagnetic order
is very unlikely to occur.

In summary we have studied the effects of nonmagnetic impurities in the 
superconducting state of the $t$-$J$ model using the Bogoliubov de Gennes 
equation derived via a slave-boson mean-field approximation. 
Near the impurity the $d$-wave OP is suppressed, 
and the extended $s$-wave component is induced as expected also from
the GL description. The CDW-like feature due to the Friedel
oscillations is most likely responsible for the double peak structure
found in the LDOS at energies close to zero. We conclude that the
violation of time reversal symmetry is suppressed here due to the
removal of a quasiparticle bound state at zero energy. This is in
contrast to the situation at the surface or at interfaces between
$d$-wave superconductors.

  We are grateful to M. Ogata, H. Fukuyama and C. Honerkamp for
helpful discussions.  
  We also thank T. Nishino for useful advice on numerical 
  calculations. 
  K.K was supported by Grant-in Aid for Scientific Research on Priority 
  Areas "Anomalous metallic state near the Mott transition" from 
  the Ministry of Education, Science and Culture of Japan.



\bigskip
\bigskip

  {\bf Fig. 1}  
  The spatial variation of $\Delta_d$ in the system with an impurity.
  Here $\delta=0.20$, $t/J=3$ and the system size is 
  $N = 16 \times 16$. 
  
  {\bf Fig. 2}
  The spatial variation of $\Delta_s$ in the system with 
  an impurity. Here $\delta=0.20$, $t/J=3$ and the system size is 
  $N = 16 \times 16$.

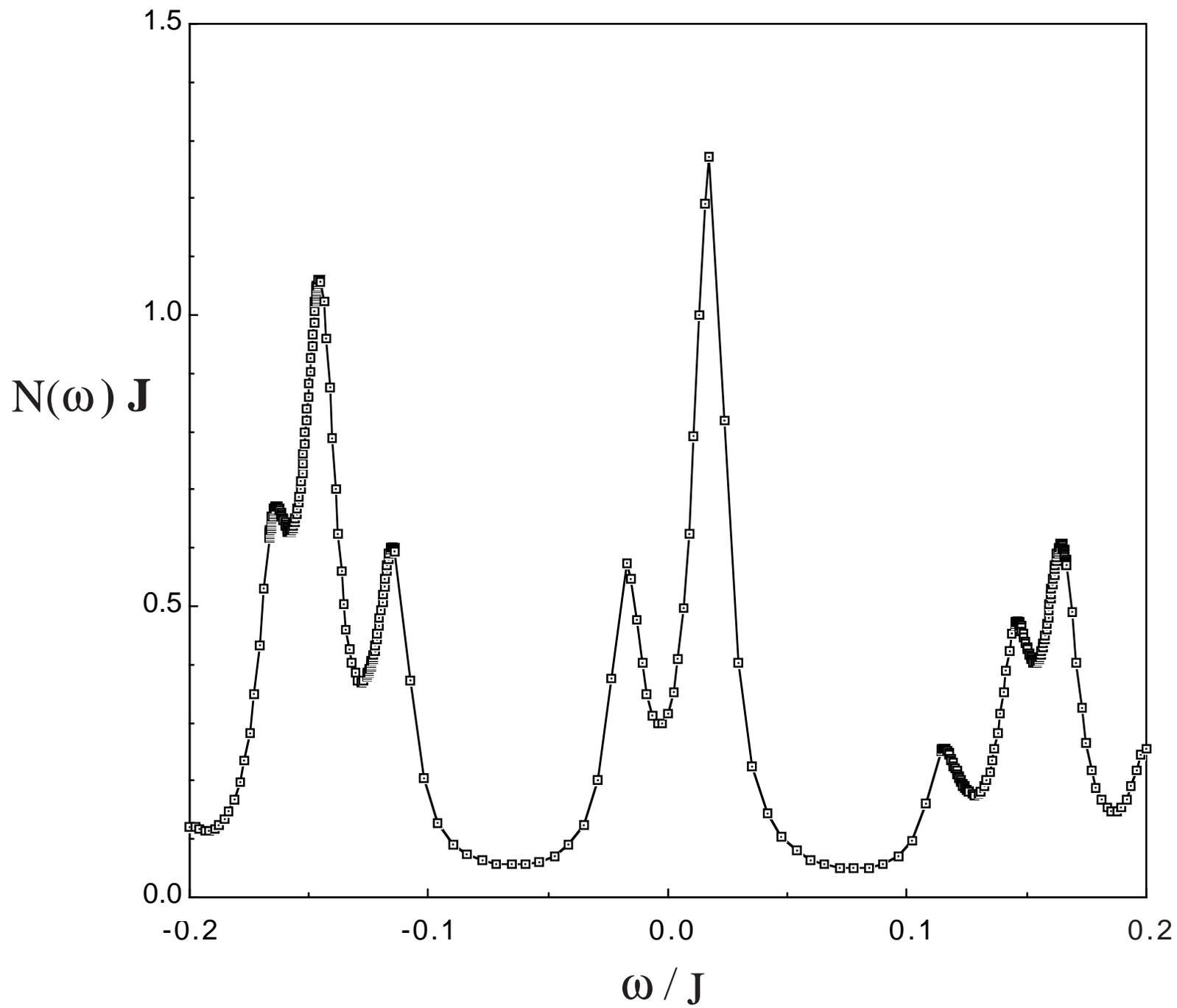
  {\bf Fig. 3} 
  LDOS at a nearest neighbor site of the impurity   
  with $\delta = 0.20$, $t/J = 3$ and $N =18 \times 18$. 
  Here we have introduced finite width $\Gamma = 0.008J$ to each state.    

  {\bf Fig. 4}   
  The electron density around the impurity. 
  Here $\delta=0.20$, $t/J=3$ and the system size is 
  $N= 18 \times 18$.


\begin{thebibliography}{99}
\bibitem{Nishi} For numerical diagonalization of matrices we have used 
the computer program written by H. Nishimori; TITPACK version2.
\end{thebibliography}
\end{document}